\documentclass[smallcondensed]{svjour3}     
\smartqed  
\usepackage{multicol}
\usepackage{graphicx}
\usepackage{subfigure}
\usepackage{algorithm}
\usepackage{algorithmic}
\usepackage{color}
\usepackage[none]{hyphenat}
\usepackage{diagbox}
\usepackage{multirow}
\usepackage{amsmath}

\begin{document}

\title{GC-LSTM: Graph Convolution Embedded LSTM for Dynamic Network Link Prediction
}


\author{Jinyin~Chen     \and Xueke Wang     \and   Xuanheng~Xu 
         }


\institute{J. Chen \at
                Institute of Cyberspace Security and the College of Information Engineering, Zhejiang University of Technology, Hangzhou, China. \\
              Tel.: +(86) 13666611145\\
              \email{chenjinyin@zjut.edu.cn}
           \and
           X. Wang \at
             the College of Information Engineering, Zhejiang University of Technology, Hangzhou, China.\\
              Tel.: +(86) 17660478061\\
              \email{3049128970@qq.com}           
           \and
           X. Xu \at
              the College of Information Engineering, Zhejiang University of Technology, Hangzhou, China. \\
              Tel.: +(86) 15757171889\\
              \email{1225910594@qq.com}           
           \and
            }

\date{Received: date / Accepted: date}

\maketitle
\begin{abstract}
\sloppy{}
Dynamic network link prediction is becoming a hot topic in network science, due to its wide applications in biology, sociology, economy and industry. However, it is a challenge since network structure evolves with time, making long-term prediction of adding/deleting links especially difficult. Inspired by the {\color{blue}{great}} success of deep learning frameworks, especially the convolution neural network (CNN) and long short-term memory (LSTM) network, {\color{blue}{we propose a novel end-to-end model with a Graph Convolution Network(GCN) embedded LSTM, named GC-LSTM, for dynamic network link prediction.}} Thereinto, LSTM is adopted as the main framework to learn the temporal features of all snapshots of a dynamic network. While for each snapshot, GCN is applied to capture the local structural properties of nodes as well as the relationship between them. One benefit is that our GC-LSTM can predict both added and removed links, making it more practical in reality, while most existing dynamic link prediction methods can only handle removed links. {\color{blue}{Extensive experiments demonstrated that GC-LSTM achieves outstanding performance and outperforms existing state-of-the-art methods.}}
\end{abstract}

\begin{multicols}{2}
\section{Introduction}
\label{sec:intro}
Network analysis has emerged as a powerful tool in various {\color{blue}{fields}}~\cite{2018Optimizing,sun2017complex,wang2017investigating,ma2020streaming}, such as social networks~\cite{2020Applications,laishram2016link} and economy~\cite{Kazemilari2015Correlation}. Most {\color{blue}{existing}} algorithms focus on the analysis of static networks~\cite{Fu2018Link,Xuan2018Social}, but ignore the dynamics of real world systems. Dynamic networks~\cite{trivedi2017know} {\color{blue}{can make up for the weakness and attract researchers' attention}}. In particular, dynamic network link prediction (DNLP) {\color{blue}{has become a hot topic in}} recent studies. DNLP aims to predict future linkage status of the given network based on historical information. For instance, we can predict people's future relationships in social networks\cite{2020Applications,chen2018fast} according to their historical behaviors, friends and {\color{blue}{even their}} personal attributes. Such examples could also be found in scientific collaboration networks~\cite{lande2020link}, social security networks~\cite{wang2020seven}, disease transmission networks~\cite{lentz2016disease}{\color{blue}{, etc.}}

\begin{figure*}[!t]
\centering
\includegraphics[width=0.8\linewidth]{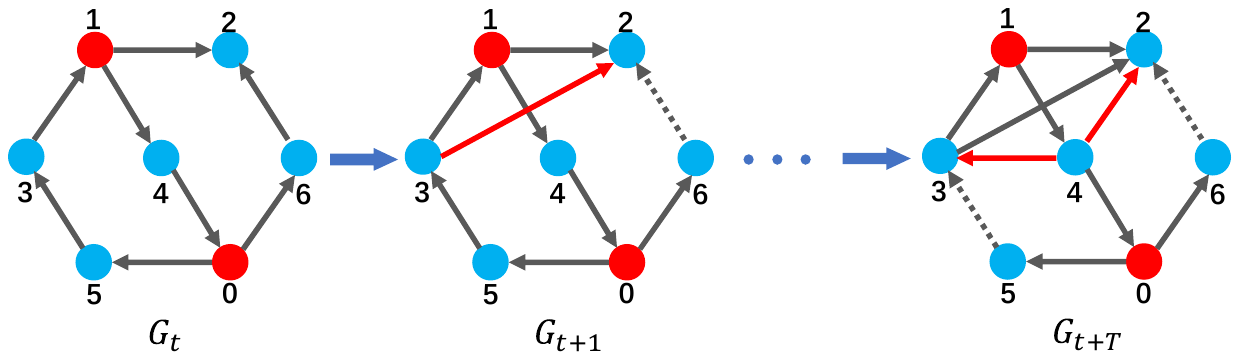}
\caption{An example of dynamic networks.}
\label{Fig1}
\end{figure*}

Given a dynamic network, it is often modeled as a chronological sequence of graphs {\color{blue}{with fixed node and changed links.}} This characterizes the dynamics of networks while static network models could not. Fig.~\ref{Fig1} shows an example of a dynamic network. The relationships between entities could last during the whole observed period (the black line), or {\color{blue}{occur at a certain moment}} (the red line), or disappear at some time (the dotted black line). {\color{blue}{Even if the different entities (the red nodes) have similar network structures at the beginning, they may evolve in different ways.}} DNLP tries to learn the {\color{blue}{evolution}} patterns of each node from historical information and then predicts the structure of next intermediate graph.

Static network link prediction approaches, including similarity {\color{blue}{indexes}} and network embedding techniques, only make prediction based on a {\color{blue}{specific}} snapshot. {\color{blue}{Though sometimes it is effective,}} they could not capture the evolution patterns due to the ignorance of temporal information. Although graph convolution network~(GCN)~\cite{kipf2017semi}, graph attention network~(GAT)~\cite{zhang2018multiresolution} and other GNNs have shown their {\color{blue}{abilities}} of handling high dimensional graph-structured data, they still {\color{blue}{suffer from}} the same problem and are limited to static network analysis. A bunch of dynamic network analysis methods, especially DNLP algorithms, have been {\color{blue}{proposed for better learning}} the temporal features. Some are developed by extending time series processing to similarity {\color{blue}{indexes}} and network embedding approaches. And many are based on recurrent neural networks~(RNNs) like long short-term memory network~(LSTM). In dynamic networks, the emerging/disappeared links {\color{blue}{are more worthy of}} attention than those existing during the whole period.

Motivated by the excellent performance of LSTM~\cite{hochreiter1997long} in handling time series data and {\color{blue}{GCN's feature learning capacity}}, we propose an unified model {\color{blue}{capable of handling the}} spatio-temporal data. Similar idea was once proposed by  Manessi et al.\cite{MANESSI2020107000}. They also jointly made use of GCN and LSTM in one model in {\color{blue}{a}} simple way of stacking GCN layer and LSTM cell sequentially, while we embed GCN into LSTM cells to better integrate structural information. However, their proposed method {\color{blue}{can not}} realize dynamic link prediction. In particular, {\color{blue}{we propose}} a novel end-to-end dynamic network link prediction deep model, called GC-LSTM, which could handle links that are going to appear or disappear. The main idea of our model is to make full use of GCN to learn network structure in the hidden state and cell state, and learn the temporal feature through LSTM model. GC-LSTM can effectively handle high-dimensional, time-dependent, and sparse structural sequence data. {\color{blue}{We conduct}} extensive experiments on real-world data sets. {\color{blue}{The results show}} that our model is significantly better than the current state-of-the-art methods. The main contributions {\color{blue}{of this paper}} are as follows.

\begin{enumerate}

\item For dynamic network link prediction, we propose a novel end-to-end deep learning model, {\color{blue}{named GC-LSTM}}, which extracts the structural feature of each snapshot network through graph convolution, and learns temporal structure through LSTM. The model can effectively learn spatio-temporal features  to make precise predictions on dynamic networks.
\item {\color{blue}{Most existing methods can only predict the added links in the network, while our}} method can predict all links that are going to appear, disappear, or constant to obtain accurate predictions of the whole dynamic network evolution.
\item We conduct extensive {\color{blue}{experiments to verify the effectiveness of GC-LSTM on dynamic networks, and compare}} with different baseline methods. {\color{blue}{It is show}} that our model significantly outperforms the current state-of-the-art methods on various metrics.
\end{enumerate}

The rest of the paper is organized as follows. In Section~\ref{sec:relatedwork}, we briefly review recent literatures related to dynamic network link prediction. Then, we give a detailed interpretation of GC-LSTM in Section~\ref{sec:Methodology}. In Section~\ref{sec:experiment}, we show all the experimental results, as well as detailed analysis. Finally, in Section~\ref{Conclusion}, we summarize the main content of the paper and conclude with future works.
\section{Related Work\label{sec:relatedwork}}

Dynamic networks are often modeled as a sequence of graphs to simplify the problem. Apparently, different snapshots play different roles in the prediction. Similarity based dynamic network link prediction is a typical one. In recent years, more and more literatures are focused on the link prediction of dynamic networks. They can be classified into similarity-based methods~\cite{zhang2020temporal,wu2020link,rahman2018dylink2vec,moradabadi2016link,shang2016evolving,chi2019link}, machine/deep learning-based methods~\cite{zhou2018dynamic,pareja2020evolvegcn,goyal2020dyngraph2vec,2019lstm,chiu2018deep}, and so on~\cite{moradabadi2017link,mangal2013link,kazemi2019time2vec}.

\textbf{\emph{Similarity based methods.}} The larger similarity{\color{blue}{,}} the more likely two nodes will be connected. A number of dynamic network~\cite{trivedi2019dyrep} link prediction technologies~\cite{wang2020seven} utilize the topology information of the network to define the similarity between pairwise nodes, named structural similarity indexes, including local indexes and global indexes. The local similarity indexes only need the neighborhood information of nodes, e.g., Common Neighbors (CN)~\cite{newman2001clustering}. It simply uses the number of common neighbors of two nodes as the index but {\color{blue}{it is}} quite effective. Other common local similarity indexes {\color{blue}{include}} Jaccard (JA), Adamic-Adar (AA), Resource allocation (RA), Hub promoted index, Hub-contracted index, and Salton index etc~\cite{wang2020seven}. While the global similarity index~\cite{wang2020seven} makes full use of the global topology information of the network, such as Katz, Random Walk, SimRank, Leicht Holme Newman Index (LHNI) and Matrix Forest Index (MFI) etc.

Further, Chen et al.~\cite{chen2016supervised} proposed a supervised network link prediction method for dynamic network {\color{blue}{by using an ensemble result}} of classifiers trained for each property. However, the optimization method is computationally expensive and is limited by the {\color{blue}{existing}} similarity index. Additionally, Wang et al.~\cite{2017Link} assumed that the ability of each node
to attract links depends not only on its structural importance, but
also on its current popularity. {\color{blue}{Under}} this assumption, they proposed an approach based on SPM and took the node popularity into consideration. Similarity-based methods often make use of the common characteristics of network topology and have good generalization. However, they are limited by the time-varying features.

\textbf{\emph{Machine/deep learning based methods.}} Besides the similarity based prediction methods, machine/deep learning is also applied to calculate the optimal similarity~\cite{kipf2016semi,chen2016supervised,chen2018fastgcn,schlichtkrull2018modeling} for precise network link prediction.
Thereinto, Tomas et al.~\cite{kipf2016semi} proposed a semi-supervised learning algorithm based on spectral convolution,
which can encode the graph structure data and features effectively and has a good performance in
the field of link prediction. Additionally, Ahmed et al.~\cite{ahmed2016sampling} believed that the most recent snapshots are of great importance and thus added a damping coefficient to each snapshot in the order of time. And they further proposed Time Series Random Walk~(TS-RW) to capture the global information~\cite{ahmed2016efficient}. Other works may learn network features in deep learning frameworks.
So, Chen et al.~\cite{chen2018fastgcn} proposed a batched {\color{blue}{training method, named FastGCN, which defines}} convolution as a form of integral transformation to solve a common memory bottleneck
in GCN. They reconstructed the sampling program of loss and gradient, which is proven by
the integration transformation of the embedded function. Numerous experiments have confirmed
that FastGCN's prediction speed is one order of magnitude faster than GNN, and it maintains
a high prediction performance.

Furthermore, Michael et al.~\cite{schlichtkrull2018modeling} proposed a new link prediction {\color{blue}{method}}, named relational-GCN (R-GCN). Regarding the standard link prediction, a good competition effect is obtained and the factorization model is enriched by R-GCN.
Following, Yang et al.~\cite{yang2019advanced} {\color{blue}{proposed a novel link prediction model, named NetworkGAN, which tackles the}} challenging temporal link prediction task~\cite{zhan2020susceptible,peng2020dynamic} efficiently. {\color{blue}{This method models the spatial and temporal features simultaneously}} in the dynamic networks via deep learning techniques.
Similarly, Rahman et al.~\cite{rahman2018dylink2vec} proposed DyLink2Vec, which learns metric embedding for node pairs to predict link states. Zhou et al.~\cite{zhou2018dynamic} proposed DynamicTriad, which models the network evolution as triadic closure process. Rather than turning a dynamic network into a discrete sequence. Nguyen~\cite{nguyen2018continuous} et al. proposed temporal random walk on the original network, which is considered as continuous-time network embedding. Further, Lei et al.~\cite{lei2019gcn} proposed a nonlinear model GCN+GAN to solve link prediction tasks in
weighted dynamic networks.

Most {\color{blue}{existing}} dynamic link prediction methods focus on a pair of nodes rather than the whole snapshot when predicting and thus may not be able to integrate global information {\color{blue}{automatically}}. By comparison, our GC-LSTM method can extract both structural feature of each snapshot network and the temporal feature to make precise prediction on dynamic networks.

\section{Methodology\label{sec:Methodology}}
In this section, we will introduce the GC-LSTM model for dynamic network link prediction. It's capable of learning both structural and temporal feature of dynamic networks for {\color{blue}{the future links addition and deletion.}}

\subsection{Problem Definition}
We will give necessary definitions, {\color{blue}{which will be used}} throughout this paper. Formally, we define dynamic networks as follows.

\newtheorem{Dynamic Networks}{\textbf{\textsc{Definition}}}
\newtheorem{network link prediction}[Dynamic Networks]{\textbf{\textsc{Definition}}}

\begin{Dynamic Networks}
\textbf{(Dynamic Networks)}
Dynamic networks can be represented as a sequence of discrete snapshots, $\{G_{1}$,$\cdots$,$G_{T}\}$, where $G_{t}=(V,E_{t},A_{t})(t \in [1,T])$ represents a network at time $t$. Let $V$ be the set of all nodes and $E_{t}$ be the temporal links within the fixed timespan [$t-\tau$,$t$]. $A_{t}$ denotes the adjacency matrix of $G_{t}$, where the element $A_{t}(i,j)$=1 if there is a link between $i$ and $j$, and $A_{t}(i,j)$=0 otherwise.
\end{Dynamic Networks}

In a static network, the network link prediction problem generally aims to predict the unobserved or future links by the current network. Static link prediction methods mainly focus on the structural feature of the network. Different from static network link prediction, dynamic network link prediction also needs to learn the temporal feature of the network according to the dynamic evolution processes of previous snapshots. Our goal is to extract the structural feature of each snapshot network through graph convolution and learn temporal structure {\color{blue}{via LSTM}}.

\begin{network link prediction}
\textbf{(Network Link Prediction in Dynamic Networks)}
Given a sequence of graphs with length $T$, $\{G_{t-T},\cdots,G_{t-1}\}$. Dynamic network link prediction is considered as a structural sequence modeling problem. It aims to learn the evolution information of the previous $T$ snapshots to predict the probability of all links at time $t$. The aim is to realize the next line edge's prediction and keep the entity unchanged. defined as
\begin{equation}
\begin{aligned}
\label{Eq1}
\hat{A}_{t}=argmaxP(A_{t}\mid A_{t-T},\cdots,A_{t-1})
\end{aligned}
\end{equation}

\end{network link prediction}
where $\{A_{t-T},\cdots,A_{t-1}\}$ represents the adjacency matrices of previous $T$ snapshots, $A_{t}$ and $\hat{A}_{t}$ are the real and predicted adjacency matrices of the snapshot at time $t$, respectively.

\begin{figure*}[!t]
\centering
\includegraphics[width=0.5\linewidth]{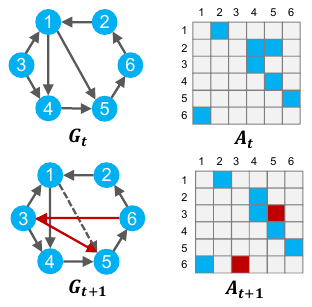}
\caption{An illustration of dynamic network evolution and its adjacency matrix. The network changes from time $t$ to $t+1$. $E(6;3$) and $E(3;5)$ emerge while $E(2;5)$ vanishes, the adjacency matrix also changes from $A_{t}$ to $A_{t+1}$ , with those elements equal to $1$ represented by filled squares.}
\label{Fig2}
\end{figure*}

Since the adjacency matrix $A_{t}$ is a precise reflection of how a network changes over time, which is adopted as the input of our prediction model, as shown in Fig.~\ref{Fig2}. We utilize a sequence of data $\{A_{t-T},\cdots,A_{t-1}\}$ with length $T$ to predict the adjacency matrix $A_{t}$, where $P(A_{t}\mid A_{t-T},\cdots,A_{t-1})$ models the probability of links to appear conditioned on the past $T$ adjacency matrices.

\begin{figure*}[!t]
\centering
\includegraphics[width=1\linewidth]{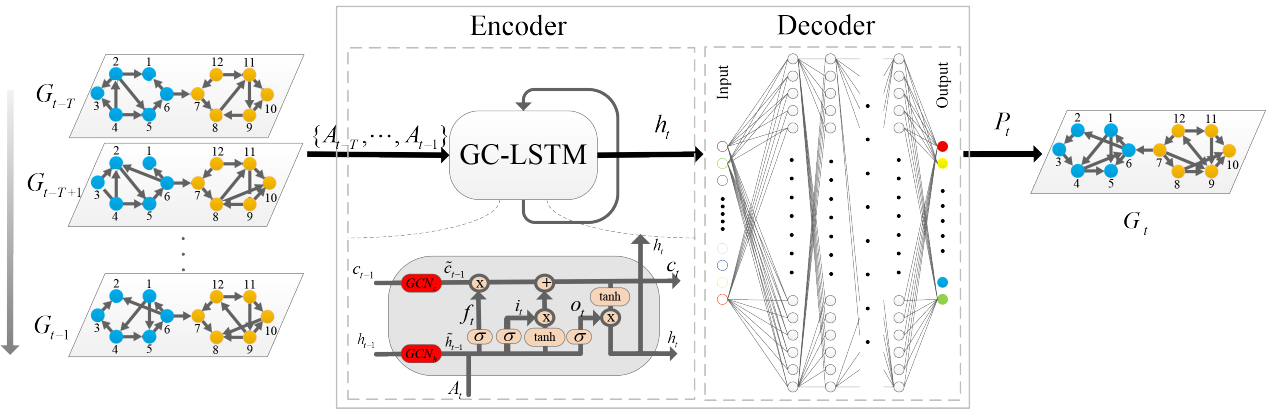}
\caption{The overall framework of encoder and decoder model for end-to-end dynamic network link prediction, in which encoder model is GC-LSTM, while decoder model is a fully connected layer network. Given a sequence of networks with length $T$, $\{G_{t-T},\cdots,G_{t-1}\}$, each network is transformed into a adjacency matrix $A_t$ as the input. The GC-LSTM is two GCNs embedded into the LSTM to learn spatio-temporal feature from the extracted data. $GCN_{h}$ and $GCN_{c}$ denote the graph convolution on long-term information $c$ and $h$, where $c_0$ and $h_0$ are zero matrixes. The decoder projects the received feature $h_{t}$ maps back to the original space to get $G_{t}$. Here, $\sigma$ in GC-LSTM is an activation function and sigmoid is adopted in this paper.}
\label{Fig3}
\end{figure*}

\subsection{Overall Framework}
The proposed GC-LSTM consists of {\color{blue}{an encoder and a decoder}} shown in Fig. \ref{Fig3}. Encoder model is GCN embedded LSTM, using GCN to learn network structure of the cell state $c$ and the hidden state $h$ of each moment snapshot, while using LSTM to learn the temporal information of the state of each link. Decoder is a fully connected layer network to convert the extracted features mapping back to the original space. GC-LSTM will output the predicted network and implement network link prediction in a unified fashion.

Before the detailed description of the proposed model, {\color{blue}{we introduce some}} terms and notations that will be used in this paper, listed in TABLE \ref{table1}. Notice that, in LSTM, subscript $f$ represents the forget gate, $i$ and $c$ represent the input gates and $o$ represents the output gate.

\begin{table*}[!ht]
\centering
\caption{Terms and notations used in GC-LSTM.}
\begin{tabular}{lr}
\hline\hline

Symbol & Definition \\ \hline
$N$ & number of nodes at each snapshot \\
$A_{t}$ & the adjacency matrix at time $t$ as the input data \\
$P_{t}$ & the output probability of each link at time $t$ \\
$d$ & number of hidden layer unit in GC-LSTM  \\
$W_{f,i,o,c}$, $b_{f,i,o,c}$ & weight and bias of three gates in GC-LSTM \\
$K$ & the order of graph convolution \\
$\sum_{k=0}^{K}\theta_{k}$ & weight of graph convolution \\
$\tilde{L}_{t}$ & the rescaled graph Laplacian matrix at time $t$ \\
${D}_{t}$ & the degree matrix of $A_{t}$ at time $t$  \\
${I}_{N}$ &  the identity matrix \\
$\lambda_{max}$ &  the largest eigenvalue of graph Laplacian matrix \\
$h_{t}$,$c_{t}$ & hidden and cell state of GC-LSTM \\
$\tilde{h}_{t}$,$\tilde{c}_{t}$ & hidden and cell state extracted by GCN \\
$W^{(l)}_{d}$,$b^{(l)}_{d}$ & weight and bias of $l^{th}$ layer of decoder \\ \hline\hline
\label{table1}
\end{tabular}
\end{table*}

\subsection{GC-LSTM Model}
We use the proposed GC-LSTM model as an encoder to extract the corresponding temporal and structural information from the structure sequence data $\{A_{t-T},\cdots,A_{t-1}\}$, and the hidden layer vector $h_{t}$ at the last moment will be used as the output of GC-LSTM.

In the task of dynamic network link prediction, the linkage status {\color{blue}{(exist or non-exist)}} of each node with others at multiple times can be regarded as a time series, equivalent to the row vector in the adjacency matrix. Generally, LSTM is applied to capture the temporal characteristics of time series data. So in GC-LSTM, we utilize the LSTM to solve long-term dependency problems and effectively learn temporal features of the dynamic graphs. The link state of each node in the dynamic network may use LSTM to implement temporal prediction, such as predicting links at the next moment. On the other hand, it is necessary {\color{blue}{to utilize the previous link state information of the nodes, and consider the impact}} of the link state of the neighbors, as well as the network structure characteristics. GCN has been proved efficient in network embedding for learning structural feature. We propose GC-LSTM model, where the Graph Convolution (GC) models are adopted to extract the structural characteristics of the snapshots at each moment, and LSTM is capable of learning temporal feature of dynamic network.

GCN is originally designed for undirected networks which means the Laplacian matrix of the symmetric network. To {\color{blue}{embed}} GCN into the proposed GC-LSTM, we first need to redefine the Laplacian matrix. Following Ma et al's idea, the directed Laplacian is defined as
\begin{small}
\begin{equation}
\begin{aligned}
    L^{sym} = I - \frac{1}{2} (\Phi^{1/2}P\Phi^{-1/2}+\Phi^{-1/2}P^T \\
     .\Phi^{1/2})
\end{aligned}
\end{equation}
\end{small}where $P=D^{-1}_{out}A$ refers to the transition probability matrix. And $\Phi=diag(\phi_{norm}(v))$ where $\phi_{norm}$ is the Perron vector of $P$ and $v$ refers to the node in $G$.
After obtaining the directed Laplacian, the graph convolution on directed networks is then defined as
\begin{equation}
    g_{\theta} = \sum_{k=0}^{K} \theta_{k}T_{k}(\tilde{L})
\end{equation}
where $\theta$ {\color{blue}{denotes}} the parameters of $g_{\theta}$ and $T_k$ is Chebyshev polynomial which is defined as $T_k(x) = 2xT_{k-1}(x)- T_{k-2}(x)$ with $T_{0}(x)=1$ and $T_{1}(x)=x$. The filter $g_{\theta}$ then becomes applicable to the analysis of directed networks. In the rest of the paper, $K^{th}$ order of graph convolution is {\color{blue}{denoted}} as $GCN^{K}(A, X)$ for simplicity.

The GC-LSTM model mainly relies on two state values, the hidden state $h$ which is used to extract the input information at the last time, and the cell state $c$ which is used to save the long-term information. The essential of GC-LSTM is that it has a cell state $c$ {\color{blue}{during}} the forward process, resulting in information {\color{blue}{transferred}} over the cell state $c$ for a long time. In the dynamic network link prediction task, we need to consider the influence of the hidden state of the neighbors on the hidden state of the node, and the influence of the cell state of the neighbors. {\color{blue}{Since the}} cell state $c$ and the hidden state $h$ respectively reflect different information. Therefore, we propose to use two GCN models to perform convolution operations on the cell layer state and the hidden layer state at the last time.

In GC-LSTM model, the first step is to decide what information will be thrown away from the previous cell state. This decision is performed by a forget gate $f_{t}\epsilon [0,1]^{d}$, where $0$ means all information will be forgotten, $1$ means all information will be reserved, defined as

\begin{figure*}[!t]
\centering
\includegraphics[width=1\linewidth]{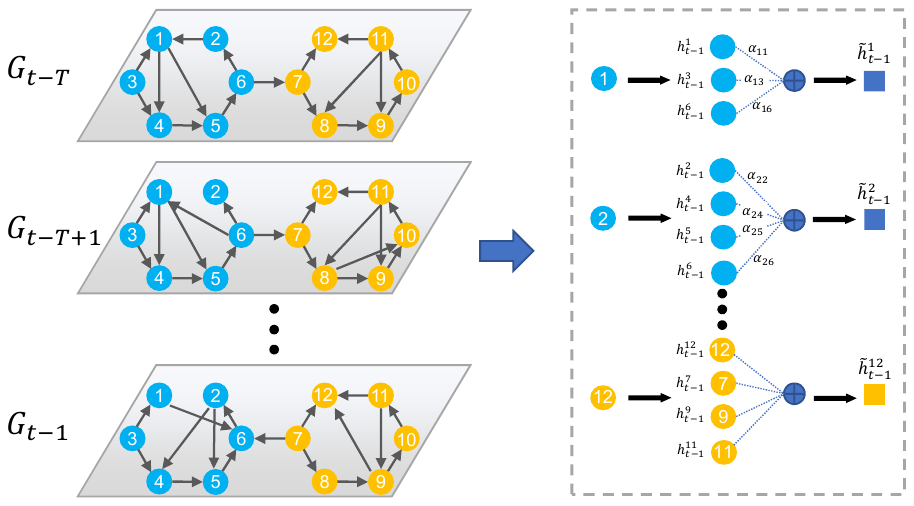}
\caption{ The framework of $1^{st}$ order GCN hidden state $h$.}
\label{FigGCN}
\end{figure*}

\begin{small}
\begin{equation}
f_{t}=\sigma (A_tW_{f}+GCN^{K}_{f}(\tilde{A}_{t-1}, h_{t-1})+b_{f})
\label{equ:ft}
\end{equation}
\end{small}where $A_{t}\epsilon R^{N\times N}$ denotes the input of the GC-LSTM at time $t$, $h_{t-1}\epsilon R^{N\times d}$  denotes the hidden state of the GC-LSTM at time $t$-1. $W_{f}\epsilon R^{N\times d}$  and $b_{f}\epsilon R^{d}$ are the weight and bias matrices of forget gate, respectively. $\sum_{k=0}^{K}\theta_{hkf}T_{k}(\tilde{L}_{t-1})h_{t-1}$ represents the graph convolution on the hidden state $h_{t-1}$ at time $t-1$. $\tilde{L}_{t-1}=\frac{2}{\lambda_{max}}L_{t}-I_{N}$ is a rescaled graph Laplacian, where $L_{t}=I_{N}-D^{-\frac{1}{2}}_{t}A_{t}D^{-\frac{1}{2}}_{t}$ is the normalized graph Laplacian. $A_{t}$ is the adjacency matrix. $D_{t}$ is the degree matrix of $A_{t}$, $I_{N}$ is the identity matrix, $\lambda_{max}$ denotes the largest eigenvalue of $L_{t}$. $\sum_{k=0}^{K}\theta_{hkf}$ are the parameter vector, and $K$ is the order of the graph convolution. $\sigma(\cdot)$ denotes the sigmoid function, $N$ denotes the input dimension, $d$ denotes the hidden layer dimension of GC-LSTM. When $K$=1, the GCN model can use the information of the $1^{st}$ order neighbors, so the operation of the $1^{st}$ GCN on the hidden state $h$ is as shown in the Fig.~\ref{FigGCN}. In addition, the operation of the GCN on the hidden state $c$ is the same as $h$, only that the input of GCN has changed from $h$ to $c$.

The next step is to update the cell state. First, a $tanh$ layer generates a new candidate vector of the cell layers, $\bar{c}_{t}\epsilon [-1,1]^{d}$. Then, $\bar{c}_{t}$ is a sigmoid layer which determines how many new candidate vector will be added to the cell state, as the input gate $i_{t}\epsilon [0,1]^{d}$. Finally, the cell state can be updated by the forget gate and the input gate, as follows.

\begin{small}
\begin{equation}
\begin{aligned}
\label{equ:ct}
&\bar{c}_{t}=tanh(A_{t}W_{c}+GCN^{K}_{o}(\tilde{A}_{t-1}, h_{t-1})+b_{c}),\\
&i_{t}=\sigma (A_{t}W_{i}+GCN^{K}_{c}(\tilde{A}_{t-1}, h_{t-1})+b_{i}),\\
&c_{t}=f_{t}\odot GCN^{K}_{c}c_{t-1}+i_{t}\cdot \bar{c}_{t}.
\end{aligned}
\end{equation}
\end{small}where $W_{i,c}\epsilon R^{N\times d}$ and $b_{i,c}\epsilon R^{d}$ are the weight and bias matrices of the input gate, respectively. $\theta_{hkc}$, $\theta_{hki}$ and $\theta_{ck}$ are the parameter vectors of the graph convolution on $h$ and $c$.

Therefore, the updated cell layer information can not only save long-term information, but also selectively filter out some useless information. Finally, we need to decide what information will be output, it is implemented by the output gate:
\begin{small}
\begin{equation}
\begin{aligned}
\label{equ:ot}
&o_{t}=\sigma (A_{t}W_{o}+GCN^{K}_{o}(\tilde{A}_{t-1}, h_{t-1})+b_{0}),\\
&h_{t}=o_{t}\odot tanh(c_{t}).
\end{aligned}
\end{equation}
\end{small}where $W_{o}\epsilon R^{N\times d}$ and $b_{o}\epsilon R^{d}$ are the weight and bias matrices of the output gate, respectively. $\theta_{hko}$ is the parameter vector of the graph convolution on $h$.

Above all, the three gates constitute the entire forward process of the GC-LSTM. In this paper, the adjacency matrices $\{A_{t-T},\cdots, A_{t-1}\}$ is used as the input data, so the hidden layer vector $h_{t}\epsilon R^{N\times d}$ obtained by GC-LSTM already contains both temporal and structural information of snapshot at the previous $T$ moments. It should be always noticed that the numbers of hidden units $d$ in GC-LSTM varies when the number of nodes $N$ is different for each dynamic network. The larger $N$ is the more units we need in the model. Generally, the number of hidden units $d$ is selected from $\{64,128,256,512\}$.

\begin{figure*}[!t]
\centering
\includegraphics[width=1\linewidth]{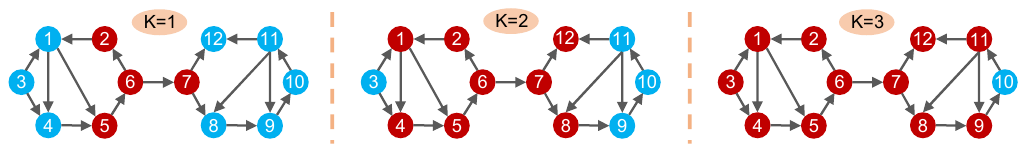}
\caption{An illustration of $K^{th}$ order (K=1, 2, 3) GCN, which uses $K^{th}$ order Chebyshev polynimial to approximate the graph convolution.}
\label{Fig4}
\end{figure*}

It is worth noticing that the GCN operation is approximated by a $K^{th}$ order Chebyshev polynomial\cite{kipf2016semi} and the hyper-parameter $K$ {\color{blue}{affects}} the performance of GC-LSTM. $K$ is associated with the richness of structural information. {\color{blue}{That means GC-LSTM can}} utilize the information with the \emph{K-hop} neighborhood of the central node. As shown in Fig. \ref{Fig4}, the information of $v_2$, $v_5$ and $v_7$ together with $v_6$ itself is considered with $K$=1. And additional information, $v_1$, $v_4$, $v_8$ and $v_{12}$, is provided when $K$=2. The larger $K$ is, the larger the neighborhood will be considered. When $K$ gets larger, the performance of GC-LSTM will be improved and it will surely increase the computational cost. To balance the performance and efficiency, we set $K$ to 3 in this paper.

\subsection{Decoder Model}
In order to output the final prediction network, we use a fully connected layer network as a decoder to turn the hidden vector $h_{t}$ at last moment in the sequence data into the final probability matrix.
\begin{equation}
\begin{aligned}
\label{equ:pt}
&Y^{(1)}_{d}=ReLU(W^{(1)}_{d}h_{t}+b^{(1)}_{d}),\\
&Y^{(k)}_{d}=ReLU(W^{(k)}_{d}Y^{(k-1)}_{d}+b^{(k)}_{d}),\\
&P_{t}=o_{t}\odot tanh(c_{t}).
\end{aligned}
\end{equation}
where $W_{d}\epsilon R^{d\times N}$  and $b_{d}\epsilon R^{N}$ are the weight and bias matrices of the decoder, respectively. $L$ denotes the number of hidden layers in the fully connected layer, and the number of units in each hidden layer may vary according to the dimension of data for better performance. $P_{t}\epsilon R^{N\times N}$ represents the final probability matrix of links, satisfying $P_{t}(i,j)\in[0,1]$ represents the probability of the link from $i$ to $j$. The lager $P_{t}(i,j)$ is, the higher probability that the two nodes $i$ and $j$ are connected.

\subsection{Loss Function and Model Training}
The purpose of training the entire GC-LSTM model is to improve the accuracy of the dynamic network link prediction, so the training is designed to make the output probability matrix more similar to the adjacency matrix at time $t$. In the regression prediction problem, the $L_2$ distance is usually used to reflect the degree of similarity between the predicted value and the true value.
\begin{small}
\begin{equation}
\begin{aligned}
L_{2}=\|P_{t}-A_{t}\|^{2}_{F}=\sum_{i=0}^{N}\sum_{j=0}^{N}(P_{t}(i,j)-A_{t}\\ .(i,j))^{2}
\label{equ:l2}
\end{aligned}
\end{equation}
\end{small}where $P_{t}$ represents the output probability matrix at time $t$, $A_{t}$ represents the real adjacency matrix at time $t$, $N$ represents the number of nodes in the snapshot at time $t$. However, if we only use the $L_2$ distance as the loss function, it will lead the predicted results biased towards zero {\color{blue}{due to the sparsity of the network.}} To avoid such overfitting to certain extent, we also employ a regularization term $L_{reg}$.

The total loss in the training process thus is defined as
\begin{small}
\begin{equation}
\begin{aligned}
\label{equ:l}
L=L_{2}+\beta L_{reg}
\end{aligned}
\end{equation}
\end{small}where $\beta$ is a parameter to trade-off the importance of $L_{2}$ and $L_{reg}$, and the best $\beta$ will be found in the model training process. The regularization loss $L_{reg}$ is calculated by calculating the sum of squares of {\color{blue}{the ownership weight}} in GC-LSTM model. In order to minimize Eq.~(\ref{equ:l}), we adopt the Adam as the optimizer of our model.

\section{Experiments\label{sec:experiment}}
The proposed GC-LSTM model is compared {\color{blue}{with the widely used baseline methods}} on several real benchmark dynamic networks to testify its performance. All experiments are implemented on 12G NVIDIA GPU, 32G DDR running memory server, which are coded by Python.

\subsection{Setup}
\textbf{Parameter Settings:} The GC-LSTM model proposed in this paper includes an encoder module composed of GCN and LSTM network. Each data set is divided according to fixed time, and 320 network sequences are taken. Among which the first 240 networks are the training set and the last 80 moments are the test set. {\color{blue}{The number of}} LSTM networks is 1. {\color{blue}{The number of}} GCN networks is 4 (corresponding to four gate units in LSTM respectively, independent {\color{blue}{with}} each other). Hyperparameter K of Chebyshev polynomial in each GCN is 3. LSTM hidden layer dimensions is 256 for the first five data sets and 512 for the last one. GCN network hidden layer dimension is 256 for the first five data sets and 512 for the last one. The iterations are 200. Eleven continuous network sequences in the training set were used as inputs to the model.

\textbf{Baseline Methods:} To validate the effectiveness of GC-LSTM, we compare it with node2vec~\cite{grover2016node2vec}, TNE~\cite{zhu2016scalable}, ctRBM~\cite{li2014deep}, GTRBM~\cite{li2018restricted}, and DDNE~\cite{li2018deep}, listed as follows.

\begin{itemize}
\item \textbf{node2vec}~\cite{grover2016node2vec}: node2vec develops a $2^{nd}$ order biased random walk procedure to explore neighborhood of a node, which can strike a balance between local properties and global properties of a network. In~\cite{nguyen2018continuous}, node2vec is adopted as a baseline for dynamic link prediction.
\item \textbf{TNE}~\cite{zhu2016scalable}: TNE models network evolution as a Markov process and then uses the matrix factorization to learn the embedding vectors for all nodes.
\item \textbf{ctRBM~\cite{li2014deep}}: ctRBM is a generative model based on temporal RBM. It first generates a vector for each node based on temporal connections and predict future linkages by integrating neighbor information.
\item \textbf{GTRBM~\cite{li2018restricted}}: GTRBM takes the advantages of both tRBM and GBDT to effectively learn the hidden dynamic patterns.
\item \textbf{DDNE~\cite{li2018deep}}: Similar to autoencoder, DDNE uses a GRU as an encoder to read historical information and decodes the concatenated embeddings of previous snapshot into
future network structure.
\end{itemize}

As the same setting in~\cite{nguyen2018continuous} for node2vec, we set the dimension of the embedded vector to 80, and find the optimal values of the hyper-parameters $p$ and $q$ from the scope $\{0.5,1,1.5,2\}$. Weighted-$L_2$~\cite{grover2016node2vec} is used to obtain an embedding vector $e^{u,v}$ for the link from node $u$ to $v$. Each element in the vector is defined as
\begin{equation}
\begin{aligned}
\label{equ:euv}
e^{u,v}_{i}=|u_{i}-v_{i}|^{2}
\end{aligned}
\end{equation}
where $u_{i}$ and $v_{i}$ are $i^{th}$ elements of embedding vectors of nodes $u$ and $v$, respectively. We compute the vector for a link by combining the learned representations of the corresponding nodes, and put the obtained representation of links into a logistic regression classifier to classify links. For TNE, we also set the dimension to 80 and regard the inner products of embedding matrix and its transpos. When we calculate the embedded vector matrix $Z$ for each node, then $Z\cdot Z^{T}$ is used to calculate the probability of all links and the threshold $\tau$=0.9 is used to judge whether the link {\color{blue}{is existing}} or not. For ctRBM and GTRBM, we set the number of visible units equal to the number of nodes in the corresponding network, and the dimension of hidden layers as 128 for the first four smaller datasets and 256 for the rest. For DDNE, we set the dimension as 128 for the first four smaller datasets and 512 for the rest.
For our proposed model, GC-LSTM, we set the hidden unit $d=256$, and the order $K$=3. In addition, we also analyze the parameter sensitivity of GC-LSTM model.

\subsection{Evaluation Metrics}
Usually, the metrics used to evaluate static network link prediction are also applied to dynamic ones, such as Area Under the ROC Curve (AUC). Based on AUC, the Area Under the Precision-Recall Curve (PRAUC)~\cite{yang2015evaluating} is designed to evaluate the sparsity of networks, and focus on the added links. Since there are also links removed in dynamic networks, Junuthula et al.~\cite{junuthula2016evaluating} restricted the measurements to only part of node pairs and proposed the Geometric Mean of AUC and PRAUC (GMAUC) for the added and removed links. Li et al.~\cite{li2014deep} used SumD to calculate the difference between the predicted network and the true one, {\color{blue}{and evaluate network link prediction}} methods in a more strict way. But the absolute difference could be misleading. For example, suppose that the SumD of the two dynamic network link prediction models are both equal to 5. If one model mispredict 5 links in a total of 10 links, while the other mispredict 5 links in a total of 100 links{\color{blue}{. It is reasonable}} to think the latter performs better than the former, while the SumD metric could not tell this.

In our experiment, at first, we choose AUC, GMAUC, and Error Rate~\cite{zhu2016scalable} as metrics to evaluate all dynamic network link prediction results.
\begin{itemize}
    \item \textbf{AUC}: If among $n$ independent comparisons, there are $n'$ times that {\color{blue}{the existing time that the existing link gets higher probability score than the non-existing}} link and $n''$ times the get the same score, then the AUC is defined as
    \begin{equation}
    \begin{aligned}
    \label{eq:AUC}
    {\rm AUC}=\frac{n'+0.5n''}{n}.
    \end{aligned}
    \end{equation}

    Before calculating the AUC index, we will select all existing links and randomly sample the same number of links that do not exist to reduce the impact of sparsity.
    \item \textbf{GMAUC}: It is a metric proposed to evaluate dynamic network link prediction, which combines botu PRAUC and AUC by taking geometric mean of the two quantities , and the GMAUC is defined as
    \begin{small}
    \begin{equation}
    \begin{aligned}
    \label{equ:GMAUC}
    {\rm GMAUC}=\Big(\frac{{\rm PRAUC}_{new}-\frac{L_{A}}{L_{A}+L_{R}}}{1-\frac{L_{A}}{L_{A}+L_{R}}} \\
    \cdot 2({\rm AUC}_{prev}-0.5) \Big)^{1/2}
    \end{aligned}
    \end{equation}
    \end{small}where $L_{A}$ and $L_{R}$ are the numbers of added and removed links{\color{blue}{, respectively,}} $PRAUC_{new}$ is the PRAUC value calculated among the new links and $AUC_{prev}$ is the AUC for the originally {\color{blue}{existing}} links.
    \item \textbf{Error Rate (ER):} It is defined as the ratio of the number of mis-predicted links $N_{false}$ to the total number of truly existent links $N_{true}$, which is defined as
    \begin{small}
    \begin{equation}
    \begin{aligned}
    \label{equ:er}
    {\rm ER}=\frac{N_{false}}{N_{true}}
    \end{aligned}
    \end{equation}
    \end{small}

    Unlike the SumD which only calculates the absolute link difference in two networks, ER considers relative link difference for precise evaluation. The smaller ER corresponds to the better prediction effect.
\end{itemize}

In particular, we further divide ER into two parts as an effective supplement, Error Rate+ (ER+) and Error Rate- (ER-). ER+ indicates the ratio of the number of mis-predicted links in truly existent links to the total number of truly existent links. ER- indicates the ratio of the number of mis-predicted links in truly non-existent links to the total number of truly existent links.

\begin{figure*}[!t]
\centering
\includegraphics[width=1\linewidth]{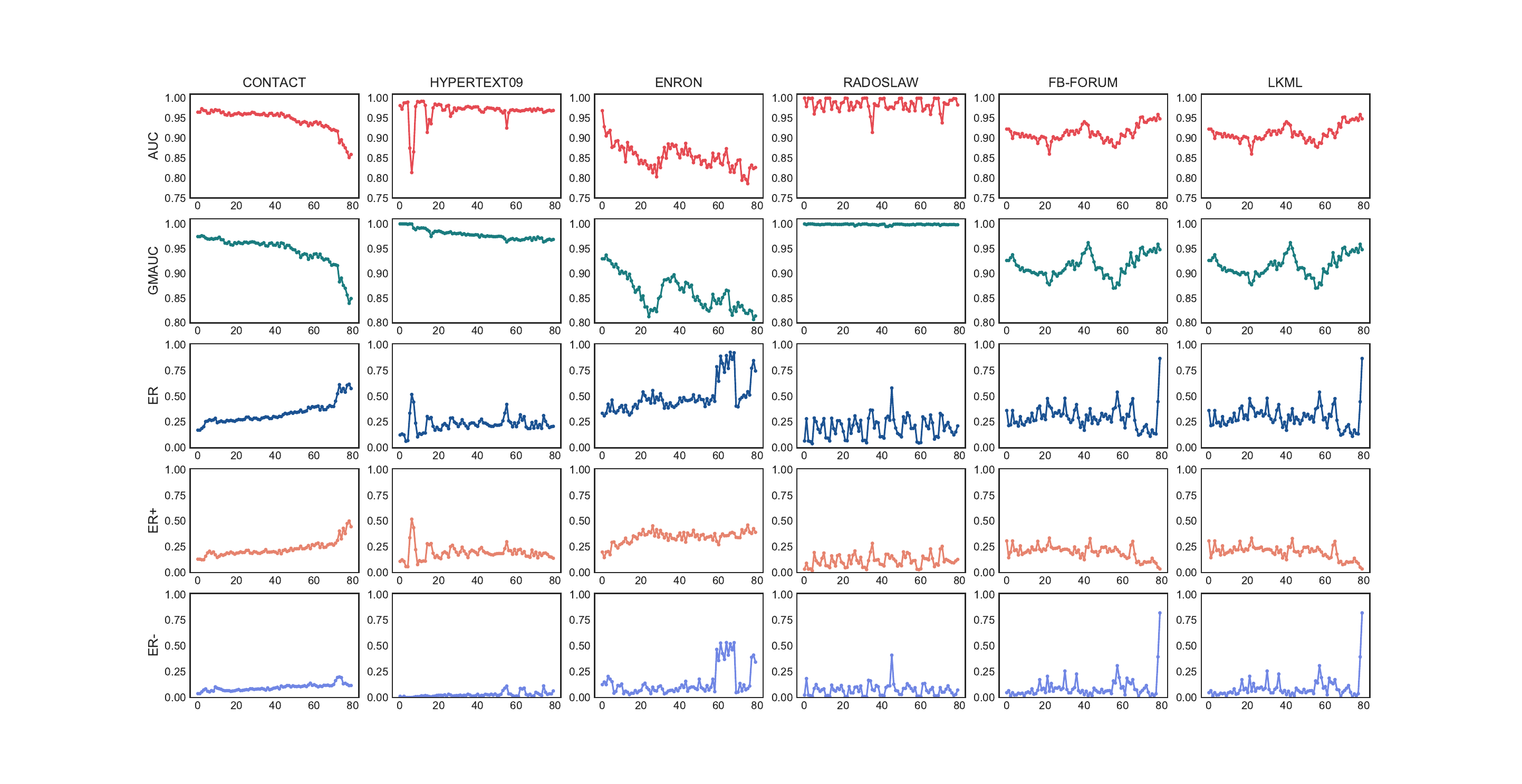}
\caption{The performance of GC-LSTM on AUC, GMAUC, ER, ER+ and ER-, obtained by our model, as functions of time $t$ for the six datasets.}
\label{Fig5}
\end{figure*}

\subsection{Datasets}
We carry out the experiments on the real-world dynamic networks. Most networks are human relation networks {\color{blue}{The nodes}} representing humans and links capturing their connections. The contacts can be face-to-face, phone, emailing etc. The data sets are described as follows.

\begin{itemize}
\item\textbf{HYPERTEXT09}~\cite{isella2011s} and \textbf{CONTACT}~\cite{chaintreau2007impact}: They are human contact dynamic networks of face-to-face proximity. The networks are collected through wireless devices carried by people who attended corresponding activity. If a person contacts with another one within a certain timestamp, a corresponding link will occur. Both of them are divided into snapshots at the time interval of 10 minutes.
\item \textbf{ENRON} and \textbf{RADOSLAW}~\cite{rossi2015network}: ENRON and RADOSLAW are email networks. Each node represents an employee of a medium-sized company, and a link occurs each time such as an email sent from one employee to another. ENRON records email data for nearly three years and RADOSLAW lasts nearly nine months. The time intervals for ENRON and RADOSLAW are 3 days and 8 hours, respectively.
\item \textbf{FB-FORUM}~\cite{opsahl2013triadic}: FB-FORUM is attained from a Facebook-like online forum of students at University of California, Irvine, in 2004. It is an online social network where nodes are users and links represent interactions (e.g., messages) between users. And the records span more than 5 months. We split FB-FORUM at the time interval of 8 hours.
\item \textbf{LKML}: It is collected from Linux kernel mailing list. In this network, the nodes represent users which are identified by their email address. And each link donates a reply from one user to another. In this paper, we only focus on the 2210 users that were recorded from 2007-01-01 to 2007-04-01, and then construct a dynamic network based on the links between these users that appear from 2007-04-01 to 2013-12-01. The time interval for LKML is 7 days.
\end{itemize}

TABLE \ref{table2} provides a summary of all dynamic networks and their statistics for evaluation. $V$ represents the number of nodes. $|E_{T}|$ shows the number of dynamic links in the network, $\bar{d}$ represents the average dynamic node degree, $d_{max}$ represents the maximum dynamic node degree. Timespan (days) is the span from the first network to the last network in sequential networks.

\begin{table*}[!ht]
\centering
\caption{The basic statistics of the dynamic networks.}
\resizebox{\linewidth}{!}{
\begin{tabular}{cccccc}
\hline \hline
Dataset & $|V|$ & $|E_{T}|$ & $\bar{d}$  & $|d_{max}|$ & Timespan~(days)
\\ \hline
HYPERTEXT09 & 113 & 20.8K & 368.5 & 1,483 & 2.46 \\
ENRON & 151 & 50.5K & 669.8 & 5,177 & 1,137.55 \\
RADOSLAW & 167 & 82.9K & 993.1 & 9,053 & 271.19 \\
CONTACT & 274 & 28.2K & 206.2 & 2,092 & 3.97 \\
FB-FORUM & 899 & 50.5K & 669.8 & 5,177   & 164.49 \\
LKML&2,210 & 422.4K& 34.6&47,995 &2,436.30 \\\hline \hline
\label{table2}
\end{tabular}}
\end{table*}

Before training, we generate snapshots for each dataset at fixed intervals, and then sort them in an ascending order of time to generate dynamic network structure sequence data. Considering that the connection between people may be temporary, we remove the links that do not reappear within 8 intervals. In order to get enough samples, we split each dataset into 331 snapshots according to an ascending order of time, then use continuous 11 snapshots as one sample, where the first 10 snapshots are adopted as input and the last snapshot network is used as output, equal to the sampling period $T$=10. Therefore, we can get a total of 320 samples. Then, we group the first 240 samples as a training set and the remaining 80 samples as a test set.

\subsection{Dynamic Prediction Results}
When training the model, we take 10 consecutive snapshots $\{G_{t-10},\cdots,G_{t-1}\}$ as a sample, and feed the model with it to predict the next snapshot $G_{t}$. There are two typical strategies to process networks: 1) using only (t-1)-th snapshot $G_{t-1}$ to predict the next snapshot $G_{t}$~\cite{yao2016link}; 2) integrating the previous 10 snapshots $\{G_{t-10},\cdots,G_{t-1}\}$ into one sample to predict the next snapshot $G_{t}$~\cite{li2018deep}. In this paper, we choose the first strategy for node2vec, while choose the second strategy for the others.

Since the evolution pattern of dynamic network may change over time, we select the average of the three evaluation metrics of the first 20 test samples and all 80 samples to reflect the short-term and long-term prediction performance of the model. All results are presented in TABLE~\ref{table3}, we compared several baseline methods, including node2vec, TNE, ctRBM, GTRBM, and DDNE. Link prediction performance indicators include AUC, GMAUC, ER, ER+ and ER-. We can find that GC-LSTM outperforms all baseline methods on ER, ER+, and ER-, while node2vec still performs the worst. However, under the data sets FB-FORUM and LKML, the indicators AUC and GMAUC are not optimal under GC-LSTM link prediction method, while TNE has better effect. Next, we analyze each indicator in detail. Where we can see that the GC-LSTM model outperforms all baseline methods for both the short-term and long-term prediction capabilities, while the performance of node2vec is worst in most cases. That is to say, network embedding methods, designed for static networks, are not suitable for dynamic network link prediction, while other dynamic prediction models perform much better due to their ability of capturing dynamic characteristics.
Our GC-LSTM model still performs better than other dynamic prediction models on all datasets in most cases, especially on the GMAUC which reflects the predictive performance of both added and removed links in dynamic networks.

\begin{figure*}[!t]
\centering
\includegraphics[width=0.6\linewidth]{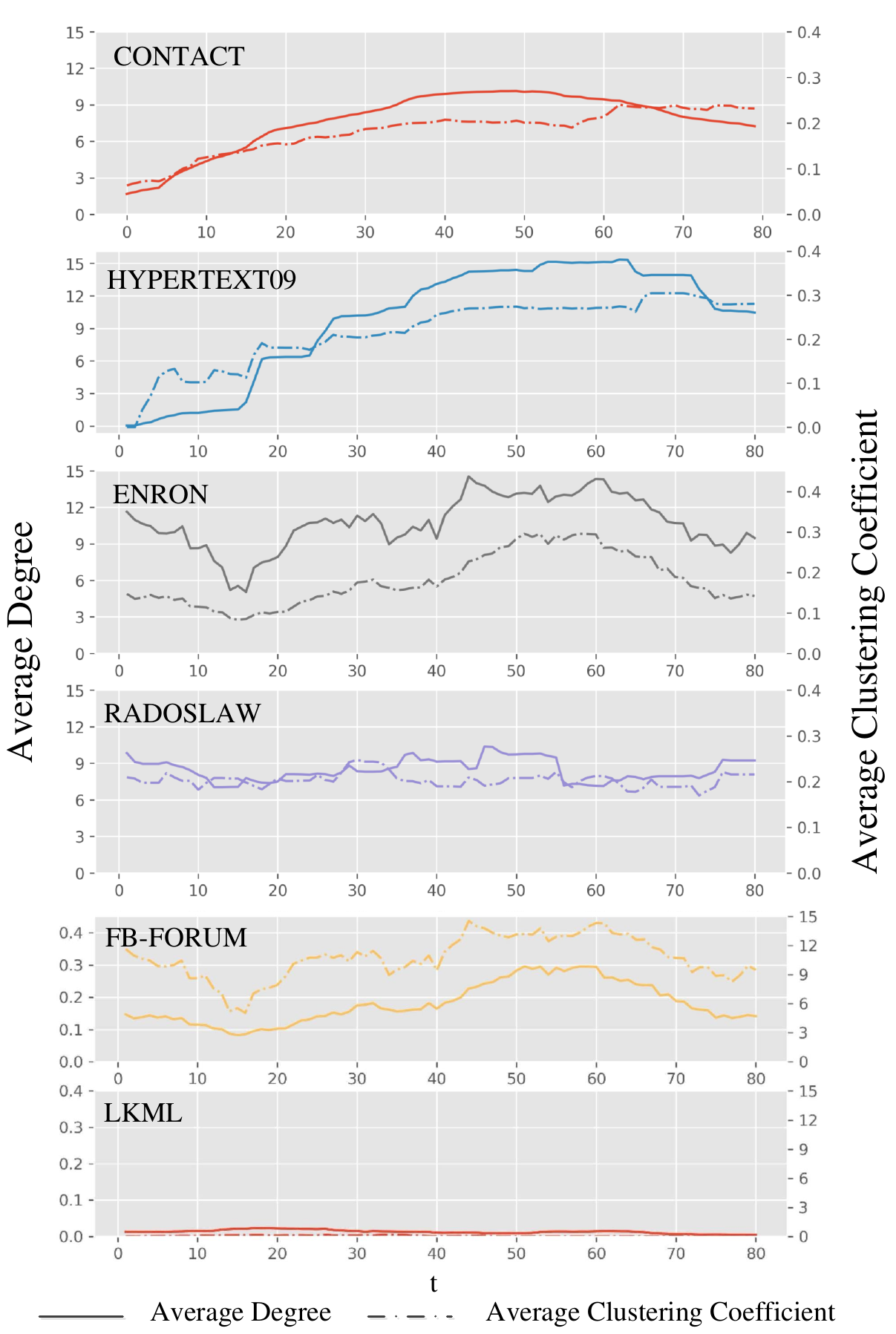}
\caption{The trends over time of network structure attributes including average degree and average cluster coefficient.}
\label{Fig6}
\end{figure*}

In addition, we also compare the difference of links between the predicted network and the ground truth network to calculate ER. The significant difference in the ER indicates that this metric is a good complement to fully measure the performance of dynamic network link prediction. As shown in TABLE~\ref{table3}, from the results of ER+ and ER-, we further find that most of baselines may predict more invalid links than the actually {\color{blue}{existed}} links, resulting in a relatively large ER. And node2vec always predict a large number of invalid links, which is 10 times or even 70 times the number of links actually exist. In particular, TNE also has a poor performance on the ER, because the matrix decomposition method used by TNE cannot balance the positive and negative samples on the sparse adjacency matrix.

\begin{table*}[!t]
\renewcommand{\arraystretch}{1.3}
\centering
\caption{Dynamic network link prediction performances on AUC, GMAUC, ER, ER+ and ER- for the first 20 samples and all the 80 samples.}\label{table3}
\resizebox{\linewidth}{!}{
\begin{tabular}{c|c|c|c|c|c|c|c|c|c|c|c|c|c}
\hline
\hline
\multirow{2}{*}{Metric} & \multirow{2}{*}{Method} & \multicolumn{2}{c|}{CONTACT} & \multicolumn{2}{c|}{HYPERTEXT09} & \multicolumn{2}{c|}{ENRON} & \multicolumn{2}{c|}{RADOSLAW} & \multicolumn{2}{c|}{FB-FORUM} & \multicolumn{2}{c}{LKML}\\ \cline{3-14}
                        &                 & 20& 80&  20& 80& 20& 80& 20& 80& 20& 80& 20& 80\\ \hline
\multirow{5}{*}{AUC}    & node2vec        & 0.5212& 0.5126& 0.6348& 0.6591& 0.7659& 0.6806& 0.6103& 0.7676&0.5142 &0.5095& 0.6348 &0.5892       \\
                        & TNE             & 0.9443& 0.9297& 0.9076& 0.8517& 0.8096& 0.8314& 0.8841& 0.8801&\textbf{0.9810} &\textbf{0.9749}& \textbf{0.9861} &\textbf{0.9867}       \\
                        & ctRBM           & 0.9385& 0.9109& 0.9029& 0.8847& 0.8468& 0.8295& 0.8834& 0.8590&0.8728 &0.8349& 0.8091 &0.7729       \\
                        & GTRBM           & 0.9451& 0.9327& 0.9204& 0.8936& 0.8527& 0.8491& 0.9237& 0.9104&0.9023 &0.8749& 0.8547 &0.8329       \\
                        & DDNE            & 0.9347& 0.9433& 0.9282& 0.9105& 0.7985& 0.7638& 0.9027& 0.8974&0.9238 &0.8729& 0.9328 &0.9115       \\ \cline{2-14}
                        & GC-LSTM         & \textbf{0.9649}& \textbf{0.9453} & \textbf{0.9573}& \textbf{0.9675}& \textbf{0.8425}& \textbf{0.8135}& \textbf{0.9838}& \textbf{0.9833} &0.9021 &0.9098 &0.9082 &0.9000       \\ \hline
\multirow{5}{*}{GMAUC}  & node2vec        & 0.1805& 0.1398& 0.4891& 0.5163& 0.4069& 0.5417& 0.7241& 0.7203&0.2744 &0.2886& 0.2309 &0.2193        \\
                        & TNE             & 0.9083& 0.8958& 0.8856& 0.8392& 0.8233& 0.7974& 0.8282& 0.8251&\textbf{0.9689} &\textbf{0.9629}& \textbf{0.9839} &\textbf{0.9778}        \\
                        & ctRBM           & 0.9126& 0.8893& 0.9139& 0.8779& 0.7207& 0.6921& 0.8004& 0.7998&0.8926 &0.8632& 0.7723 &0.7206        \\
                        & GTRBM           & 0.9240& 0.9136& 0.9407& 0.9201& 0.9148& 0.8675& 0.9157& 0.8849&0.9329 &0.9117& 0.6529 &0.6038       \\
                        & DDNE            & 0.8925& 0.8684& 0.9484& 0.9127& 0.8724& 0.8476& 0.8938& 0.8724&0.9126 &0.9023& 0.7894 &0.7809        \\\cline{2-14}
                        & GC-LSTM         & \textbf{0.9680}& \textbf{0.9456}& \textbf{0.9921}& \textbf{0.9788}& \textbf{0.8571} & \textbf{0.8134} &\textbf{0.9980}& \textbf{0.9983} &0.9062 &0.9123 &0.9184 &0.9049    \\ \hline
\multirow{5}{*}{ER}     & node2vec        & 44.7753& 25.2278& 24.3980& 12.8260& 23.9053& 24.8060& 20.7240& 21.2489&40.5109 &48.5376& 53.2895 &61.0274        \\
                        & TNE             & 13.1410& 7.1556& 1.6453& 1.9675& 23.1276& 19.9167& 16.7078& 16.7175&19.1058 &24.4350& 18.5702 &18.2091        \\
                        & ctRBM           & 0.9126& 0.8893& 0.6663& 0.6970& 2.4890& 2.7328& 1.8920& 2.0937&3.4509 &3.6782& 2.9903 &3.3089  \\
                        & GTRBM           & 0.9240& 0.9136& 0.5745& 0.5992& 1.5947& 1.8836& 1.9079& 2.0031&2.2347 &2.4396& 2.5351 &2.7942   \\
                        & DDNE            & 0.8925& 0.8684& 0.5820& 0.6084& 1.7664& 1.9014& 1.6316& 1.5941&1.9014 &1.8266& 2.0134 &2.2258   \\\cline{2-14}
                        & GC-LSTM         & \textbf{0.2324}& \textbf{0.3213}& \textbf{0.1988}& \textbf{0.2194}& \textbf{0.3763}& \textbf{0.3928} & \textbf{0.1783}  & \textbf{0.1891}&\textbf{0.2702}&\textbf{0.3012}&\textbf{0.7369}&\textbf{0.7798}
        \\ \hline
\multirow{5}{*}{ER+}    & node2vec        & 5.0931& 1.3049& 3.0721& 2.0933& 4.9831& 7.1045& 2.8204& 3.1249& 7.8192& 9.1033& 6.3719 & 13.0938\\
                        & TNE             & 3.2049& 3.2307& 0.4201& 0.2049& 3.2924& 3.9837& 3.2145& 3.6537& 7.0203& 7.9924& 4.3001 & 7.9342 \\
                        & ctRBM           & 0.2909& 0.3334& 0.1896& 0.1263& 0.6037& 0.9026& 0.4722& 0.6777& 0.9764& 1.1120& 0.9024 & 1.0017 \\
                        & GTRBM           & 0.3029& 0.2198& 0.1308& \textbf{0.1038}& 0.4908& \textbf{0.3296} & 0.3790& 0.3802& 0.9022& 0.9982& 0.8517 & 1.3826\\
                        & DDNE            & 0.2132& 0.2205& \textbf{0.1253}& 0.1104& 0.3923& 0.4828& 0.3001& 0.3841& 0.6437& 0.8029 & 0.8429 & 0.9997\\ \cline{2-14}
                        & GC-LSTM         & \textbf{0.1703}& \textbf{0.2304} & 0.1902& 0.1923& \textbf{0.3070}& 0.3763& \textbf{0.1008}  & \textbf{0.1167}&\textbf{0.2186} &\textbf{0.1997} &\textbf{0.6198} &\textbf{0.6459}       \\ \hline
\multirow{5}{*}{ER-}    & node2vec        & 39.6822& 23.9229& 21.3259& 10.7327& 18.9222& 17.7015& 17.9036& 18.124& 32.6917& 39.4343& 46.9176 & 47.9336 \\
                        & TNE             & 9.9361& 3.9249& 1.2252& 1.7626& 19.8352& 15.933& 13.4933& 13.0638& 12.0855& 16.4426& 14.2701 & 10.2749    \\
                        & ctRBM           & 0.6217& 0.5559& 0.4767& 0.5707& 1.8853& 1.8302& 1.4198& 1.4160& 2.4745& 2.5662& 2.0879 & 2.3072  \\
                        & GTRBM           & 0.6211& 0.6938& 0.4437& 0.4954& 1.1039& 1.5540& 1.5289& 1.6229& 1.3325& 1.4414& 1.6834 & 1.4116  \\
                        & DDNE            & 0.6793& 0.6479& 0.4567& 0.4980& 1.3741& 1.4186& 1.3315& 1.2100& 1.2577& 1.0237& 1.1705 & 1.2261 \\ \cline{2-14}
                        & GC-LSTM         & \textbf{0.0621}& \textbf{0.0908} &\textbf{0.0084} & \textbf{0.0270}& \textbf{0.0857}& \textbf{0.1418}& \textbf{0.0704} & \textbf{0.0724}  & \textbf{0.0515}   &\textbf{0.1014}&\textbf{0.1171}&\textbf{0.1339}    \\ \hline
                        \hline
\end{tabular}}
\end{table*}

Actually, the results prove once again that our GC-LSTM model has better performance in dynamic network link prediction. Because the baselines directly predict the next snapshot network based on the first 10 snapshot networks, while our method requires pre-training models to learn the evolution pattern of dynamic networks more effectively from the first 240 samples.
And GC-LSTM model not only uses LSTM to learn the temporal characteristics of the sequence network, but also uses GCN to learn the network characteristics of each snapshot. Therefore, in the test process, it can predict future links more {\color{blue}{accurately with lower ER}}.

Interestingly, in TABLE~\ref{table3}, the ER- is much larger than ER+ in most cases, especially for node2vec, indicating that most of the methods are more likely to predict non-existent links than existent ones. This phenomenon may be due to the fact that most of the dynamic networks in the experiments are relatively sparse.
For our GC-LSTM method, although ER+ is slightly larger than ER-, both of its ER+ and ER- are smaller than that of baselines in most cases. That is, GC-LSTM method not only has relatively high prediction accuracy on the {\color{blue}{non-existing}} links, but also has better prediction performance on the {\color{blue}{existing}} ones, i.e., it can better predict the dynamic links of the network. In general, the link prediction method proposed in this paper has better prediction effect.

Next, for test samples $G_{t}$ ($t$ changes from 1 to 80), we plot the AUC, GMAUC, ER, ER+, and ER- metric curves over time obtained by our GC-LSTM model on all networks to reflect the dynamic network link prediction performance, as shown in Fig.~\ref{Fig5}. As time evolves, AUC and GMUUC curves are gradually decreasing, while the ER, ER+, and ER- curves are gradually increasing. This phenomenon indicates that long-term link prediction for dynamic network is relatively difficult due to the uncertainty of dynamic network evolution over a long period of time. Interestingly, since the network structure of RADOSLAW and LKML evolves periodically, the performance of GC-LSTM on RADOSLAW and LKML are relatively stable. In order to further explain this phenomenon, we studied the trend of the two most common structural properties (i.e, the average degree and the average clustering coefficient) of the dynamic networks with increasing time $t$, and the results are shown in Fig.~\ref{Fig6}. We can see that the average degree and average clustering coefficients of the four datasets, CONTACT, HYPERTEXT09, ENRON and FB-FORUM, have changed significantly over time, while those of RADOSLAW and LKML are relatively stable. These results explain why we can get better long-term predictive performance.

\subsection{Prediction of Important Links}
In summary, although some methods have superior performance in the statistical metrics AUC or GMAUC, they are not satisfactory in terms of ER and predict many wrong invalid links. In some real-world scenarios, we may focus more on whether the most important links are predicted correctly. Therefore, we further evaluate the performance of all models on some important links. Here, we use two metrics to measure the importance of each link: degree centrality (DC) and edge intermediate centrality (EBC). DC is generally used to measure the node's importance according to the number of neighbors. In this paper, we use the sum of the two nodes' DC to measure the importance of the link between them. We select the top 10\% important links based on DC and EBC, and then calculate the ER for these important links, as shown in TABLE \ref{table4}. In most cases, our GC-LSTM model has the lowest ER on all data sets for short-term and long-term prediction. This validates that our GC-LSTM model still performs quite well in predicting the important links. Furthermore, comparing the results of TABLE \ref{table3} and TABLE \ref{table4}, we find that the ER on the top 10\% important links is much smaller than that on all links. This indicates that the performance of our model on those important links are better than those on less important ones.

\begin{table*}[!ht]
\renewcommand{\arraystretch}{1.3}
\centering
\caption{Prediction ER of the top 10\% important links in terms of DC and EBC.}
\label{table4}
\resizebox{\linewidth}{!}{
\begin{tabular}{c|c|c|c|c|c|c|c|c|c|c|c|c|c}
\hline\hline
\multirow{2}{*}{Metric} & \multirow{2}{*}{Method} & \multicolumn{2}{c|}{CONTACT} & \multicolumn{2}{c|}{HYPERTEXT09} & \multicolumn{2}{c|}{ENRON} & \multicolumn{2}{c|}{RADOSLAW}& \multicolumn{2}{c|}{FB-FORUM} & \multicolumn{2}{c}{LKML}\\ \cline{3-14}
                        &                & 20& 80& 20& 80& 20& 80& 20& 80& 20& 80& 20& 80\\ \hline
\multirow{5}{*}{DC}     & node2vec       & 0.6279& 0.6297& 0.4397& 0.4837& 0.4900& 0.4524& 0.4735& 0.5203&0.3873 &0.3454&0.5034&0.5289  \\
                        & TNE            & 0.9622& 0.9551& 0.2869& 0.3360& 0.3446& 0.3315& 0.5068& 0.4413&0.0595 &0.0558&0.6390&0.6288   \\
                        & ctRBM          & 0.2739& 0.3307& 0.1842& 0.2105& 0.4193& 0.4410& 0.3028& 0.3097&0.1095 &0.1137&0.3291&0.3341   \\
                        & GTRBM          & 0.2209& 0.2390& 0.2073& 0.2377& 0.4098& 0.4322& 0.2109& 0.2198&0.1127 &0.1239&0.2973&0.3030   \\
                        & DDNE           & 0.1293& 0.1359& 0.1096& 0.1147& 0.2270& 0.2133& 0.0803& 0.1249&0.1190 &0.1088&\textbf{0.1653}&\textbf{0.1821}    \\\cline{2-14}
                        & GC-LSTM        & \textbf{0.1679}& \textbf{0.1422}& \textbf{0.0438}& \textbf{0.0663}& \textbf{0.1173}  & \textbf{0.1455}   & \textbf{0.0289}    & \textbf{0.0357}&\textbf{0.0043} &\textbf{0.0075} &0.2006 &0.2955       \\ \hline
\multirow{5}{*}{EBC}    & node2vec       & 0.6747& 0.6509& 0.5602& 0.5064& 0.4607& 0.5953& 0.4657& 0.4397&0.6517 &0.6799&0.8729&0.8698    \\
                        & TNE            & 0.9998& 0.9987& 0.6655& 0.4705& 0.9598& 0.9590& 1.0000& 1.0000&1.0000 &0.9986&1.0000&0.9992    \\
                        & ctRBM          & 0.5396& 0.5619& 0.4998& 0.5203& 0.6512& 0.7381& 0.2165& 0.2291&0.4432 &0.4508&0.7279&0.7503    \\
                        & GTRBM          & 0.4418& 0.4573& 0.4040& 0.4324& 0.6906& 0.7420& 0.2399& 0.2511&0.4507 &0.4529&0.6370&0.6524    \\
                        & DDNE           & \textbf{0.2713}& \textbf{0.2849}& 0.3019& 0.3157& 0.4988& 0.5471& 0.2083& 0.2508&0.2697 &0.3014&0.6435&0.6614    \\\cline{2-14}
                        & GC-LSTM        & 0.3252 & 0.4551 & \textbf{0.1948} & \textbf{0.2615}& \textbf{0.3584} &\textbf{0.3985}& \textbf{0.1455}& \textbf{0.1588}&\textbf{0.1694} &\textbf{0.1591} &\textbf{0.4975} &\textbf{0.6038} \\ \hline\hline
\end{tabular}
}
\end{table*}

\section{Conclusion\label{Conclusion}}
In this study, we propose a new deep learning model, GC-LSTM, as an encoder-decoder architecture for dynamic network link prediction. The entire GC-LSTM model consists of LSTM and GCN, where LSTM is used to learn the temporal characteristics from continuous snapshots, while GCN is used to learn the structural characteristics of the snapshot at each moment. A fully connected layer network is used as a decoder to convert the extracted spatio-temporal features back to the original space. The proposed model can capture not only the time dependence between a sequence of snapshots, but also the impact of the network structure, i.e., it can better understand the network evolution pattern. Finally, we conducted many experiments to compare our GC-LSTM model with traditional network link prediction methods on various dynamic network datasets. The results validate that our model outperforms the others in terms of AUC, GMUUC and ER. Meanwhile, we also find that GC-LSTM model behaves well in predicting those important links characterized by DC and EBC. {\color{blue}{It is better than the others in most cases.}}

Since GC-LSTM needs pre-training, its time complexity is relatively high. In the future, we will focus on accelerating the model and predicting dynamic link for large-scale dynamic networks which may consist of hundreds of thousands nodes and links, to make it more practical in reality.

\section*{Acknowledgements}
This research was supported by the National Natural Science Foundation of China under Grant No. 62072406,
the Natural Science Foundation of Zhejiang Province under Grant No. LY19F020025,the Major Special Funding for Science and Technology Innovation 2025 in Ningbo under Grant No. 2018B10063,
the National Key Research and Development Program of China under Grant No. 2018AAA0100801,
the Special Technology Project for Pre-research of Military Common Information System Equipment under Grant No. JZX6Y201907010467.
\\

\noindent
\textbf{Compliance with Ethical Standards}

\textbf{Ethical approval:} This article does not contain any studies with human participants or animals performed by any of the authors.

\textbf{Conflict of Interest:} The authors  declare that they have no conflict of interest.
\end{multicols}


\bibliographystyle{unsrt}
\bibliography{mybibfile}

\end{document}